\documentclass[12pt]{article}
\usepackage{epsfig}
\usepackage{psfrag}
\usepackage{amsmath}
\usepackage{lscape}
\usepackage{chemarr}
\usepackage{multirow}
\usepackage{graphics}
\usepackage[super, comma, sort&compress]{natbib}
\newcommand{\bfig}{\begin{figure}[h] \begin{center}}
\newcommand{\efig}{\end{center} \end{figure}}

\begin{document}

\begin{center}
Symmetry of forward and reverse path populations
\end{center}

\begin{center}
Divesh Bhatt, Daniel M. Zuckerman\footnote{email: ddmmzz@pitt.edu}
\end{center}
\begin{center}
Department of Computational Biology,
University of Pittsburgh
\end{center}

{\bf Abstract}

 {\it In this note, we address formally the issue of symmetry of different pathways
in the forward and reverse directions in general case beyond equilibrium, making
clear the conditions necessary.
We also clarify when the symmetry is to be expected from a practical point
of view.}

We address the issue of symmetry of different pathways in the forward and 
reverse directions at steady state
formally and exactly, making clear the conditions necessary 
for a strict symmetry of relative pathway populations.  From a practical, 
approximate point of view, we will see that when user--defined 
states correspond to physical basins of reasonable depth, then path symmetry 
is expected.  Our discussion builds on prior work by Crooks\cite{crooks}
and by vanden Eijnden.\cite{eijnden} 

We begin by considering a situation of equilibrium as sketched in
Figure~\ref{sym_equil}, constructed from a 
large ensemble of systems undergoing natural dynamics.  For simplicity, 
we assume there are two stable states (A and B) and two distinct pathways 
or channels ($i$ and $j$) connecting the states as shown in 
Figure~\ref{sym_equil}, 
but our discussion is more generally applicable. In equilibrium, 
both the probability density and probability flows are unchanging in time, 
reflecting averages over the large ensemble. 

The ensemble can be usefully decomposed in several ways. First, if 
we consider a single point in time, each system in the ensemble either 
is in one of 
the states (A or B) or not.  We will focus on the fraction of systems 
not in either state, which can be further classified if we assume complete 
knowledge of the past and future of each system.\cite{eijnden}
In particular, all systems 
which were most recently in state A will proceed either back to state A or 
make a transition to state B.  A similar description applies to systems 
most recently in B, leading to the schematic flows shown in Figure~\ref{sym_equil}
These classifications apply for arbitrary definitions of states A and B 
–- whether physically reasonable or not. 

We now want to consider the relative probabilities of two pathways or 
``channels'' (which can be arbitrarily defined) such those schematized 
by $i$ and $j$ in Figure~\ref{sym_equil}.  In equilibrium, there is 
no net flow anywhere in configuration space.  For instance, along the 
surface shown as a straight bar across channel $i$ in 
Figure~\ref{sym_equil}, there will 
be an equal number of forward and reverse-moving trajectories in the ensemble.  
This balance within a pathway, in turn, requires
 that the relative probability of the two channels in the 
A--to--B direction must be matched exactly by that in the reverse 
B--to--A direction.  If the relative probabilities were different, 
a net flow would occur, violating equilibrium.
Thus, the symmetry relation is established for arbitrary definitions of
A and B, in equilibrium.

Is the symmetry of the ratio of path probabilities expected in simulations
that are not run at equilibrium? The answer depends on the conditions
and the state definitions. Exact symmetry, regardless of the states,
is expected if simulations are run in one of the special 
steady states schematized in 
Figure~\ref{sym_ab} that constitute an exact decomposition 
of equilibrium.  In particular, for the A--to--B direction, if trajectories 
arriving to B are fed back into A exactly as they would arrive there in 
equilibrium then the relative
probability of the $i$ to $j$ channels in A--to--B direction 
will remain unchanged from equilibrium.  
(Equivalently, symmetry will hold if the trajectories entering back into 
state A are distributed such that the probability distribution within 
state A is always the equilibrium probability distribution.)
Analogous considerations 
apply for the B--to--A direction (see Figure~\ref{sym_ab} (b)), 
establishing that the symmetry 
of the $i$ to $j$ population ratio is expected in such simulations 
regardless of the state definitions. 

When feedback schemes for establishing steady states do not exactly 
replicate equilibrium –- or when transition trajectories are generated 
outside the rubric of a steady state (as in the present report) -– 
we can inquire whether the symmetry condition might hold approximately.  
In other words, under what conditions are the precise details of the 
feedback scheme (or the scheme for initializing trajectories) 
unimportant?  Such insensitivity should occur if the user--defined 
A and B states are ``reasonably deep'' physical basins 
of attraction.  Here, ``deep'' means that trajectories which enter 
the state are likely to remain there long enough to explore the basin 
fully and emerge in a quasi-Markovian way -– i.e., to emerge the way 
trajectories would in equilibrium regardless of where they entered. 
Said another way, approximate symmetry is expected when intra--state
timescales are much less than transition times.

In complex systems, such as that considered here, it may be difficult 
to define true physical basins, so the forward--reverse symmetry relation 
of channel probabilities should be viewed as an approximate guideline 
or reference point.

\clearpage

\bfig
\resizebox{3.5in}{!}{\includegraphics{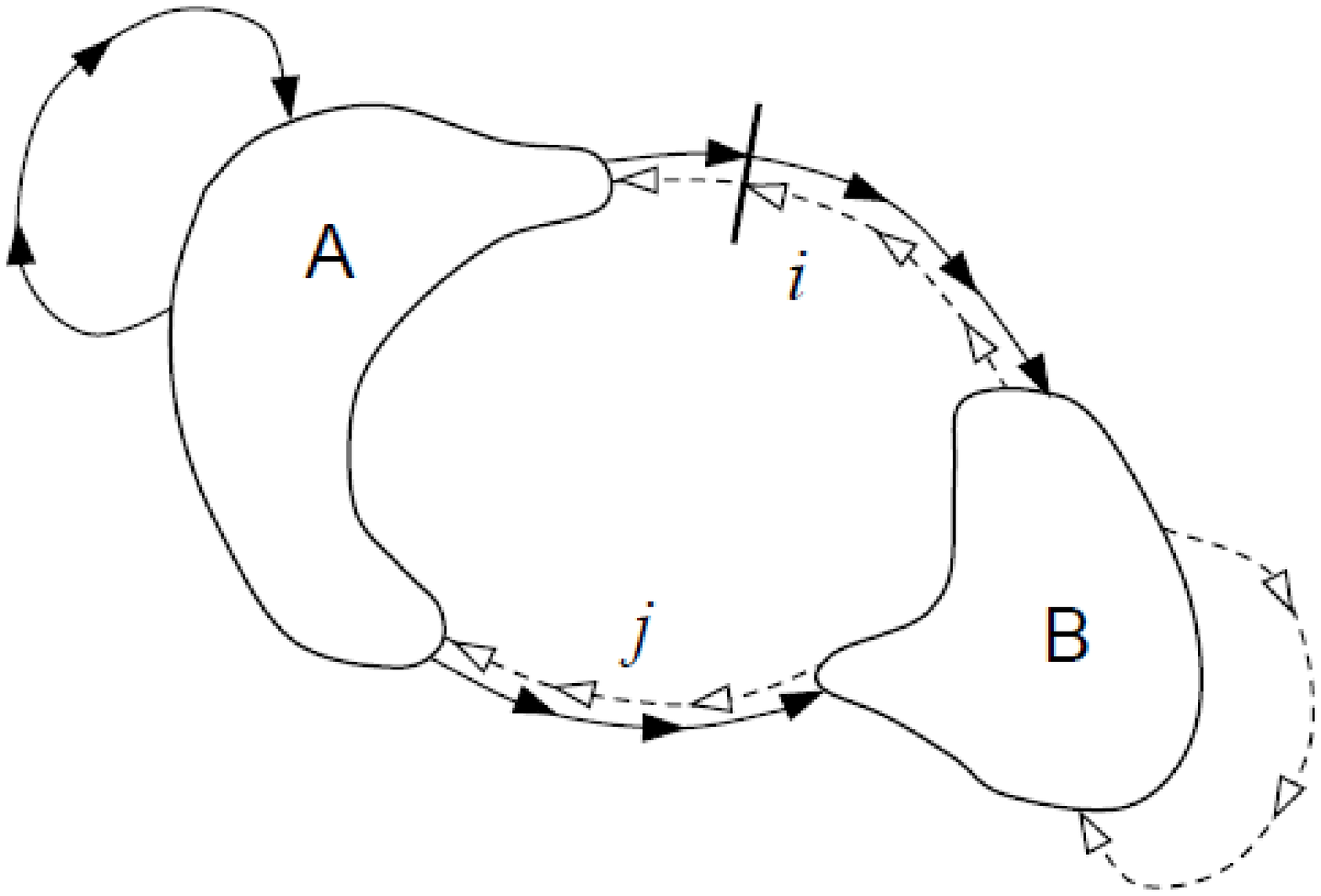}}
\caption{Schematic depiction of a system at equilibrium with two states, 
A and B.
Transitions between the two states occur via two distinct paths, $i$ and
$j$. Directed lines are used to classify possible paths: trajectories 
starting from
A that reach B before coming back to A or come back to 
A before reaching B (solid lines), 
and trajectories starting from B that
either reach A before coming back to B or come back to 
B before reaching A (dashed lines). 
At equilibrium, the net flux across any surface (such
as the solid bar across path $i$) is zero.}
\label{sym_equil}
\efig

\clearpage

\bfig
\begin{tabular}{c}
\resizebox{3.5in}{!}{\includegraphics{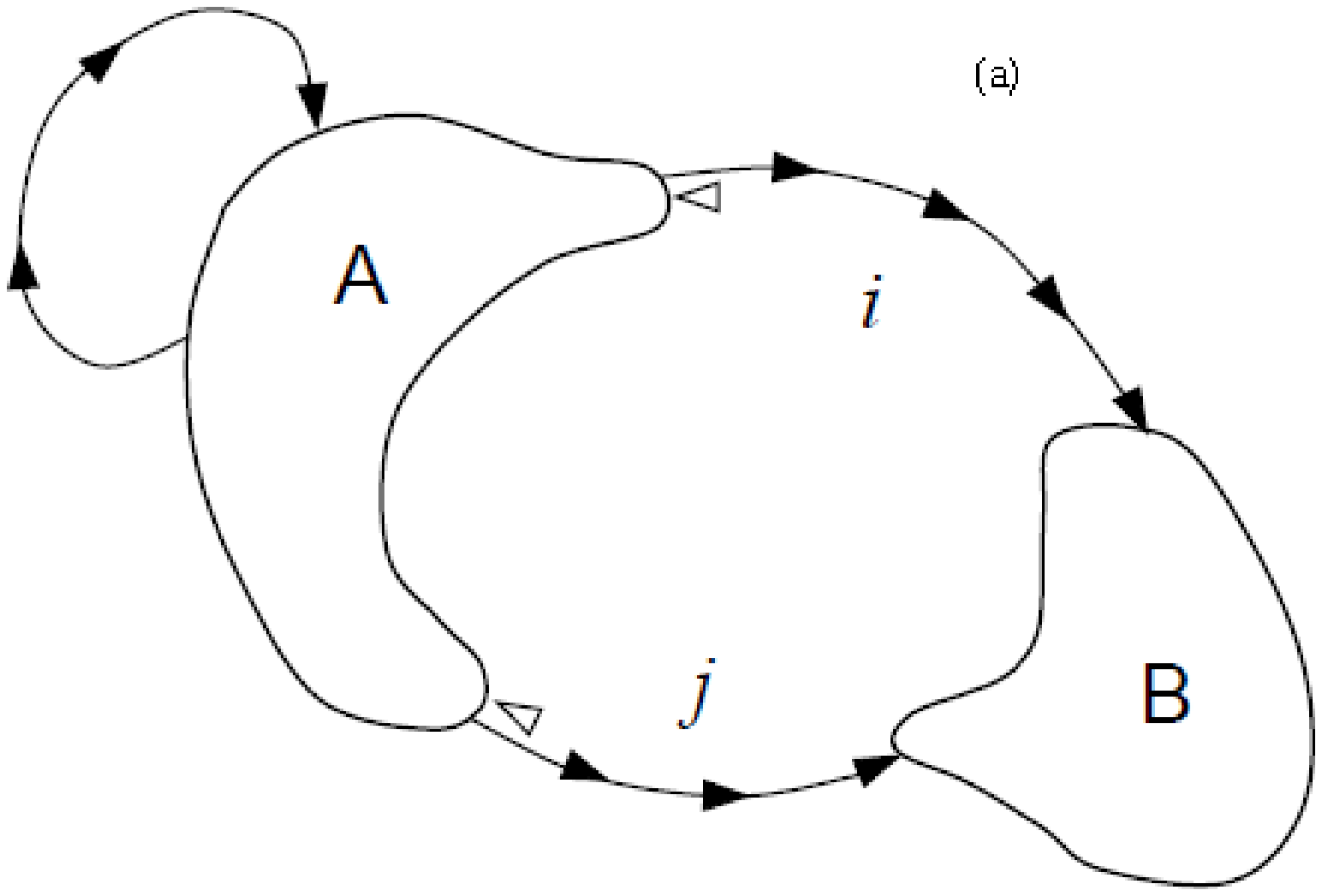}}\\
\resizebox{3.5in}{!}{\includegraphics{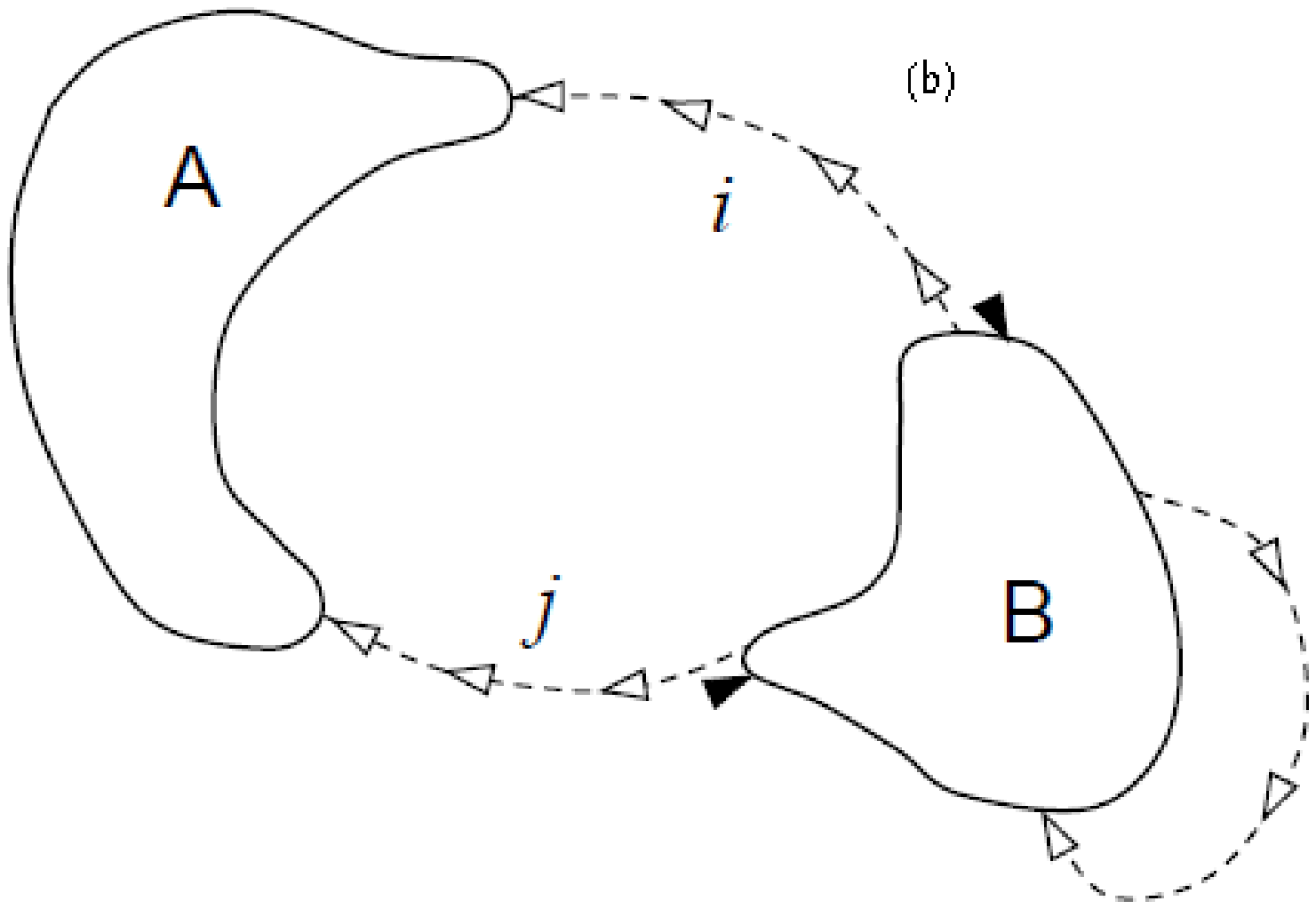}}\\
\end{tabular}
\caption{An exact decomposition of equilibrium (Figure~\ref{sym_equil})
into opposing steady states.
Panel (a) shows feedback into A of trajectories that reach state B,
and (b) feedback into B of trajectories that reach state A. 
In panel (a), solid lines and filled arrows show the trajectories 
starting from A, whereas dashed lines and 
open arrows illustrate the trajectories fed back into A from the two 
channels
$i$ and $j$. The feedback is such that the trajectories are fed back into A in
exactly the same manner as at equilibrium. A similar description is applicable
for the reverse transition in (b).}
\label{sym_ab}
\efig

\end{document}